\documentclass[aps,prb,twocolumn,amssymb,amsmath,superscriptaddress]{revtex4-1}
\usepackage{bm}
\usepackage{times}
\usepackage{graphicx}
\usepackage{color}
\usepackage{dcolumn}
\usepackage[colorlinks=true, letterpaper=true, pdfstartview=FitV, linkcolor=blue, citecolor=blue, urlcolor=blue]{hyperref}
\usepackage{appendix}
\usepackage[normalem]{ulem}

\begin{document}
\title{Universal behaviors of magnon-mediated spin transport in disordered nonmagnetic metal-ferromagnetic insulator heterostructures}
\author{Gaoyang Li}
\affiliation{College of Physics and Optoelectronic Engineering, Shenzhen University, Shenzhen 518060, China}
\author{Fuming Xu}
\affiliation{College of Physics and Optoelectronic Engineering, Shenzhen University, Shenzhen 518060, China}
\author{Jian Wang}
\email[]{jianwang@hku.hk}
\affiliation{College of Physics and Optoelectronic Engineering, Shenzhen University, Shenzhen 518060, China}
\affiliation{Department of Physics, The University of Hong Kong, Pokfulam Road, Hong Kong, China}

\begin{abstract}
We numerically investigate magnon-mediated spin transport through nonmagnetic metal/ferromagnetic insulator (NM/FI) heterostructures in the presence of Anderson disorder, and discover universal behaviors of the spin conductance in both one-dimensional (1D) and 2D systems. In the localized regime, the variance of logarithmic spin conductance $\sigma^2(\ln G_T)$ shows a universal linear scaling with its average $\langle {\rm ln}G_{T}\rangle$, independent of Fermi energy, temperature, and system size in both 1D and 2D cases. In 2D, the competition between disorder-enhanced density of states at the NM/FI interface and disorder-suppressed spin transport leads to a non-monotonic dependence of average spin conductance on the disorder strength. As a result, in the metallic regime, average spin conductance is enhanced by disorder, and a new linear scaling between spin conductance fluctuation rms($G_T$) and average spin conductance $\langle G_{T}\rangle$ is revealed which is universal at large system width. These universal scaling behaviors suggest that spin transport mediated by magnon in disordered 2D NM/FI systems belongs to a new universality class, different from that of charge conductance in 2D normal metal systems.
\end{abstract}

\maketitle

\section{introduction}

At low temperature, quantum interference in mesoscopic transport leads to universal fluctuation of charge conductance,\cite{Umbach,Stone1,Altshuler,Lee0,Qiao10,Lei14,Han19} whose magnitude depends only on the dimensionality and symmetry of the system. On the other hand, disorder-induced destructive interference can transform a metal into an insulator, known as Anderson localization.\cite{AL} The single-parameter scaling (SPS) theory\cite{Edwards,Abrahams2,Anderson,Kramer,Abrahams,Mott,Muller,Shapiro1,Shapiro2} was proposed to interpret Anderson localization in disordered systems. The scaling behavior of charge conductance needs to be considered in terms of its distribution function.\cite{Anderson} The SPS theory states that the conductance distribution has a universal form, which is determined by a single parameter, the ratio of system size $L$ to the localization length $\xi$. The localization length is obtained from the average logarithmic conductance ${\rm ln}G$ while increasing $L$: $1/\xi = -\lim _{L \rightarrow \infty} {\langle\ln G\rangle} / {2L}$.\cite{Mackinnon,Mackinnon2} Therefore, $\langle{\rm ln}G\rangle$ is widely used for verifying SPS in the localized regime.\cite{Prior,Magna,Magna2}

Statistical properties of charge transport in one-dimensional (1D) normal systems have been thoroughly studied.\cite{Roberts,Deych1,Muttalib,Mello,Plerou,Garcia} It was found that the distribution of ${\rm ln} G$, $P({\rm ln} G)$ is Gaussian in localization limit, which is determined by the average and variance of ${\rm ln}G$, i.e., $\langle{\rm ln}G\rangle$ and $\sigma^{2}({\rm ln}G) = \langle {\rm ln^{2}}G\rangle - \langle {\rm ln}G\rangle^{2}$. A universal relation is established between them\cite{Anderson,Deych1}
\begin{equation}\label{eq:scaling}
\sigma^{2}({\rm ln}G) = A \langle {-\rm ln}G\rangle^{n} + B,
\end{equation}
where the exponent $n=1$. Eq.~(\ref{eq:scaling}) reduces the two parameters in $P({\rm ln} G)$ to one and justifies SPS in 1D systems. The situation in 2D is more complicated. Most numerical investigations are in agreement with SPS,\cite{Mackinnon,Mackinnon2,Schreiber,Prior,Somoza,Somoza2,Magna2} although some deviations were reported.\cite{Kantelhardt,Queiroz,Magna} The universal relation Eq.(\ref{eq:scaling}) was numerically confirmed\cite{Prior} in a large region from diffusive to localized regimes with $n=2/3$.\cite{note2} The distribution of ${\rm ln}G$ was found to approach Tracy-Widom distribution in the localized limit.\cite{Prior,Somoza} $n=1$ was found in disordered graphene nanoribbons (quasi-1D systems),\cite{Magna} but Eq.~(\ref{eq:scaling}) is not universal in the entire energy spectrum.

Besides charge transport, magnon-based spin transport of insulator spintronics has attracted great interest recently.\cite{Brataas2} In magnetic insulators, band structure does not allow electron transport, but pure spin current can be carried by a collective mode called spin wave or magnon. Experimentally, magnon spin current is generated via spin pumping\cite{Tserkovnyak2,Tserkovnyak3,Azevedo,Saitoh} driven by ferromagnetic resonance or spin Seebeck effect (SSE)\cite{Uchida,Uchida2,Bauer,Adachi,Tang,Wu,Miao} and detected in another nonmagnetic metal (NM) via inverse spin Hall effect. The most popular structure for SSE is the bilayer structure Pt/YIG. Pt is a heavy metal with strong spin-orbit coupling, and YIG is a ferromagnetic insulator (FI) with long propagating distance of spin wave. Anderson localization of magnon spin transport in 1D magnetic insulators has been reported recently,\cite{Jakobsen,Yang} where disorder always suppresses spin transport.

For nonmagnetic metal/ferromagnetic insulator (NM/FI) heterostructures, a theoretical formalism based on the non-equilibrium Green's function (NEGF) has been developed to study the magnon-mediated spin transport properties.\cite{Li} NEGF is suitable for describing mesoscopic transport phenomena, including thermal or phonon transport.\cite{JSWang14,CZhang21,ZZYu21,JSWang23} Up to now, investigation on the statistical properties of spin transport mediated by magnon in NM/FI systems is still absent. In this work, we numerically investigate magnon-mediated spin transport in disordered 1D and 2D NM/FI systems in different transport regimes. In 2D system, new universal statistical behaviors of spin conductance are discovered in the diffusive and localized regimes, which are different from that of charge conductance in normal systems. The distribution of spin conductance as well as its higher order cumulants are also studied.

The rest of this paper is organized as follows. Section \ref{sec:model} introduces the model Hamiltonian and the NEGF method for spin transport in NM/FI heterostructures. Numerical results and relevant discussion are presented in Sec. \ref{sec:res}. Finally, a conclusion is shown in Sec. \ref{sec:conclusion}.

\section{MODEL AND THEORETICAL FORMALISM}\label{sec:model}

\begin{figure}
\includegraphics[width=7cm]{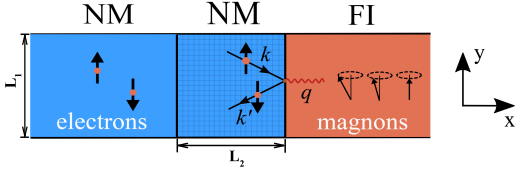}%
\caption{\label{fig:setup}Schematic view of the 2D NM/FI system. The left NM lead and the right FI lead are connected to the central NM scattering region, which is of width $L_1$ and length $L_2$. Magnon-mediated transport is along the $x$ direction. A typical experimental setup for this model is the bilayer structure of Pt/YIG.}
\end{figure}

In this section, we introduce the Keldysh theory for spin transport. The system under investigation is shown in Fig.~\ref{fig:setup}, whose Hamiltonian is
\begin{equation}
H = H_0 + H'. \label{eq:H}
\end{equation}
$H_{0}$ is the unperturbed Hamiltonian, and $H'$ is the perturbative coupling. $H_{0}$ consists of three parts: the left NM lead, the central NM region, and the right ferromagnetic insulator lead,
\begin{equation}
H_{0} = H_L + H_d + H_R.
\end{equation}
The left lead is described by noninteracting electrons,
\begin{equation}
H_L= \sum_{k\sigma}( \varepsilon_{k\sigma}-\mu_{L\sigma}) c_{k\sigma}^{\dag}c_{k\sigma},
\end{equation}
where $\mu_{L\sigma}$ is the chemical potential for spin $\sigma=\{\uparrow, \downarrow\}$. The right FI lead is described by the Heisenberg model $H_{R}=-J\sum_{<ij>}[\frac{1}{2}S_{i}^{+}S_{j}^{-}+\frac{1}{2} S_{i}^{-}S_{j}^{+}+ S_{i}^{z}S_{j}^{z}]$, where $S_{i}^{\pm}$ is the raising (lowering) operator for the localized spin at site $i$, $S_{i}^{z}$ is the spin operator in $z$ direction at site $i$, and $J$ is the exchange coupling. At low temperature, it can be approximated by free magnons\cite{HP} ($\hbar=1$):
\begin{equation}
H_R \simeq \sum_{q} \omega_q a^{\dagger}_{q} a_{q},
\end{equation}
where $a_{q}^{\dag} (a_{q})$ creates (annihilates) a magnon with momentum $q$ and $\omega_{q}$ is the magnon dispersion dependent on material details. The central region Hamiltonian is expressed as
\begin{equation}
H_{d}=\sum_{n \sigma} \epsilon_{n \sigma} d_{n \sigma}^{\dagger} d_{n \sigma}.
\end{equation}
In numerical calculations, Anderson-type disorder as random on-site potentials is added in the central region.
The perturbative coupling $H'$ has two parts,
\begin{equation}
H' = H_{T} + H_{sd}.
\end{equation}
$H_{T}$ is the coupling between the left lead and the central region,
\begin{equation}
H_T=\sum_{k\sigma n} \left[t_{k\sigma n} c^{\dagger}_{k\sigma} d_{n\sigma} + t_{k\sigma n}^{*} d^{\dagger}_{n\sigma} c_{k\sigma}\right].
\end{equation}
Following Ref.~[\onlinecite{Zheng}], the electron-magnon coupling between the central region and the right FI lead is described by the $sd$-type exchange interaction,
\begin{equation}
H_{sd}= -\sum_{q nn'} J_{qnn'} \left[ d^{\dagger}_{n\uparrow} d_{n'\downarrow}a^{\dagger}_{q} + d^{\dagger}_{n'\downarrow} d_{n\uparrow}a_{q} \right],
\end{equation}
which describes a magnon-mediated spin-flipping scattering between $n$ and $n'$ states, by absorbing or emitting a magnon with momentum $q$. $J_{qnn'}$ is the scattering strength, assumed to be weak and treated perturbatively in the NEGF method. There is an extra factor $\sqrt{2S_0}$ associated with $a^\dagger_q$ and $a_q$ in $H_{sd}$ which we dropped for the moment, where $S_0$ is the length of lattice spin. We have neglected inelastic processes such as the electron-phonon interaction and other relaxation mechanisms and therefore focus on spin current at low temperature.

When a temperature gradient $\Delta T$ is applied across the NM/FI interface, a pure spin current is generated by the spin Seebeck effect. Due to current conservation, the spin current flowing through this NM/FI heterostructure equals that flowing in the right FI lead,\cite{J-Ren} which is given by
\begin{equation}
I_{s} = i\sum_{qnn'} J_{qnn'}[ \langle  d^{\dagger}_{n\uparrow} d_{n'\downarrow}a^{\dagger}_{q}\rangle -\langle a_{q} d^{\dagger}_{n'\downarrow} d_{n\uparrow}\rangle]. \label{spin1}
\end{equation}
In DC case, the spin current in this system is obtained in the Keldysh theory as\cite{Li}
\begin{equation}
I_{s}= -i\int \frac{dE}{2\pi} {\mathrm Tr}[D_{L\uparrow}(E)({\bar \Sigma}_{R\uparrow}^<(E)-2f_{L\uparrow}(E){\mathrm Im}{\bar \Sigma}_{R\uparrow}^a(E))], \label{eq:isr}
\end{equation}
where $D_{L\sigma}=G_{\sigma}^r \Gamma_{L\sigma}G_{\sigma}^a$ is the local density of states (LDOS) matrix of electrons coming from the left lead.\cite{Buttiker} The electron Green's function is expressed as
\begin{equation}
G^r_{\sigma}=1/[g_{d\sigma}^{-1}-\Sigma_{L\sigma}^r-{\bar \Sigma}_{R\sigma}^r]= 1/[G_{L\sigma}^{-1}-{\bar \Sigma}_{R\sigma}^r].
\end{equation}
Here $\Sigma_{L\sigma}^r$ is the electron self-energy of the left lead, and $\Gamma_{L\sigma}=2\mathrm{Im}\Sigma_{L\sigma}^{a}$ is the corresponding linewidth function. In the Born approximation (BA), the effective self-energy of the right lead is ${\bar \Sigma}^<_{R\uparrow} (t,t')= iG^<_{L\downarrow}(t,t'){\Sigma}_{R}^>(t',t)$. Then the DC spin current in BA is shown as\cite{Li}
\begin{eqnarray}
I_{s} &&= -\int \frac{d\omega}{2\pi} (f^B_R(\omega)-f^B_L(\omega))\int \frac{dE}{2\pi} (f_{L\uparrow}(E)\nonumber \\
&&-f_{L\downarrow}(E+\omega)) {\mathrm Tr}[A_R(E,\omega)], \label{spin4}
\end{eqnarray}
with
\begin{eqnarray}
A_R(E,\omega)=D_{L\uparrow}(E) D^0_{L\downarrow}({\bar E})\Gamma_R(\omega), \label{ar}
\end{eqnarray}
in which ${\bar E}= E+\omega$. $f_R^B$ is the Bose-Einstein distribution for the right lead, and $f_{L\uparrow,\downarrow}$ is the Fermi-Dirac distribution for the left lead. $f_{L}^{B}$ is the effective Bose-Einstein distribution for the left lead, which is defined as $f_{L}^{B}(\omega)= 1/[e^{\beta_{L}\left(\omega+\Delta \mu_{s}\right)}-1]$. $\beta_{L}=1/k_{B}T_{L}$ is the inverse temperature of the left lead, and $\Delta\mu_{s}=\mu_{L\uparrow}-\mu_{L\downarrow}$ is the spin bias applied. $D^0_{L\downarrow}=G_{L\downarrow}^r \Gamma_{L\downarrow}G_{L\downarrow}^a$ is the partial LDOS matrix when the central region is connected only with the left lead. Note that $D_{L\uparrow}$ and $D_{L\uparrow}^{0}$ are defined in terms of different Green's functions. $D_{L\uparrow}$ is related to $G_{\sigma}$, which is the Green's function of the central region connected to both left and right leads; while $D_{L\uparrow}^{0}$ is determined by $G_{L\sigma}$, which is the Green's function when the central region is only connected to the left lead.

For simplicity, we assume that the $sd$ interaction occurs only at the same site of the NM/FI interface, which leads to the simplification $J_{qnn'} = J_{qn}\delta_{nn'}$. Without loss of generality, the hybridization function between the central region and the magnonic reservoir is assumed to be Ohmic and the linewidth function of the right lead is expressed as $\Gamma_{R}(\omega)=\pi\alpha t\omega e^{-\omega / \omega_{c}}$.\cite{J-Ren} Here $\alpha$ is the effective coupling strength, $\omega_{c}$ is the cutoff frequency, and $t$ is the hopping constant. Then the effective self-energy is formulated as
\begin{eqnarray}
\bar{\Sigma}^r_{R\uparrow}(E)  =\int \frac{d\omega}{2\pi} &&[f_R(\omega) G^r_{L\downarrow}({\bar{E}}) \nonumber \\
&&+ i f_{L\downarrow}({\bar{E}}){\mathrm Im} G^r_{L\downarrow} ({\bar{E}}) ] \Gamma_{R}(\omega), \label{sigr}
\end{eqnarray}
which is an energy convolution of local partial density of states (DOS) with the spectral function of the FI lead.

\section{NUMERICAL RESULTS AND DISCUSSION}\label{sec:res}

In this section, we investigate spin transport in disordered mesoscopic systems where quantum interference manifests. In normal metallic systems without disorder, the conductance, proportional to the total transmission coefficient of conducting channels, measures the ability of transporting electrons. For magnon-mediated spin transport in our system, the right lead is an insulator, and there is no concept of transmission. The spin conductance $G_T$ is used to measure spin transport instead. In linear response regime, spin current $I_s$ is driven by a small temperature gradient $\Delta T$, then $G_{T}=I_{s}/\Delta T$.

The Hamiltonian Eq.(\ref{eq:H}) is defined in momentum space. We can transform it to real space using the finite difference procedure.\cite{Datta} A square lattice with lattice spacing $a=5$ nm is used in tight-binding calculation, corresponding to a hopping constant $t=21.768$ meV. When Anderson disorder is present, on-site random potentials with uniform distribution $[-W/2,W/2]$ are added to the Hamiltonian of the central region, where $W$ is the disorder strength in unit of $t$. More than 10000 disorder samples are collected in numerical calculations. All calculations are carried out in the Born approximation using Eq.(\ref{spin4}). Spin conductance $G_{T}$ is in unit of $\mu$eV/K. We present numerical results on the spin transport and scaling properties of disordered 1D and 2D NM/FI systems.

\subsection{Spin transport and scaling in 1D NM/FI system}

\begin{figure}
\includegraphics[width=8.5cm]{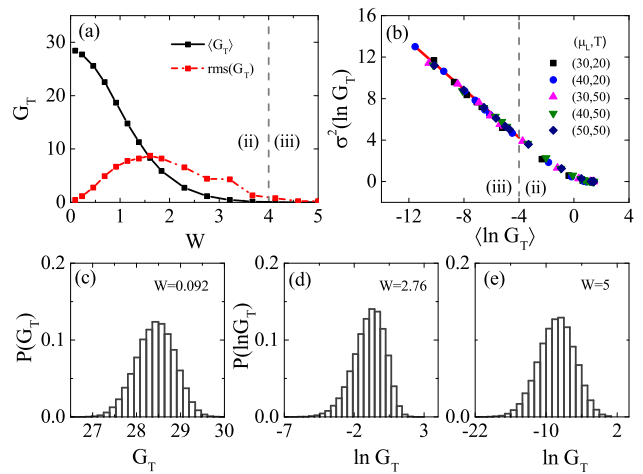}%
\caption{\label{fig:1D-G} (a) Average spin conductance $\langle G_{T}\rangle$ and its fluctuation $\mathrm{rms}(G_{T})$ as a function of the disorder strength $W$. (b) The variance of ${\rm ln}G_{T}$ dependence on $\langle {\rm ln}G_{T}\rangle$ for different Fermi energies and temperatures. (c)--(e) Spin conductance distribution $P(G_T)$ and $P(\mathrm{ln}G_T)$ for different $W$. 40000 disorder samples are collected. Parameters: $\mu_{L}=40$ meV and $T=50$ K. The gray dashed line separates the diffusive (ii) and localized (iii) regimes.}
\end{figure}

For the 1D hybrid system, we choose a central region of $1\times 30$ sites and connect it with two semi-infinite 1D leads. The cutoff frequency is fixed at $\omega_c=15$ meV, and $\alpha=0.1$. In Fig.~\ref{fig:1D-G}(a), the disorder-averaged spin conductance $\langle G_T\rangle$ and its fluctuation $\mathrm{rms}(G_T)$ are depicted as a function of $W$ for the Fermi energy $\mu_L=40$ meV and temperature $T=50$ K. Spin conductance fluctuation is defined as
\begin{equation}
{\rm rms}(G_{T}) = \sqrt{\langle G_{T}^{2}\rangle - \langle G_{T}\rangle^{2}},
\end{equation}
in which $\langle \cdots \rangle$ denotes the average over different disorder configurations for the same disorder strength $W$. Fig.~\ref{fig:1D-G}(a) shows that $\langle G_T\rangle$ and $\mathrm{rms}(G_T)$ are in the same order of magnitude, similar to charge transport in normal metal systems. The curves of $\langle G_T\rangle$ and $\mathrm{rms}(G_T)$ versus $W$ are also similar to charge conductance fluctuation in normal metal systems, which allows to identify different transport regimes: (i) metallic regime for small disorder $W$; (ii) diffusive regime centered around $W \sim 1.5$; (iii) localized regime for $W \ge 4$. The diffusive and localized regimes are separated by gray dashed lines in Fig.~\ref{fig:1D-G}(a)--(b).

The SPS for 1D normal metal systems predicts a universal exponent $n=1$ and Gaussian distribution of ${\rm ln}G$ in the localized regime. We show in Fig.~\ref{fig:1D-G}(b) the variance $\sigma^{2}({\rm ln}G_T)$ versus $\langle {\rm ln}G_{T}\rangle$ for different Fermi energies, disorder strengths, and temperatures. Clearly, all data points collapse into one single line in the localized regime. This line is fitted as
\begin{equation}
\sigma^{2}({\rm ln}G_T)=-1.16\langle {\rm ln}G_{T}\rangle - 0.38,
\end{equation}
where a universal exponent $n=1$ is identified. Distributions of spin conductance $G_{T}$ and ${\rm ln}G_{T}$ are shown in Fig.~\ref{fig:1D-G}(c)--(e) for different disorder strengths. We see that in the metallic regime with $W=0.092$, $P(G_T)$ is Gaussian like. As disorder increases, the Gaussian distribution evolves into an asymmetric non-Gaussian form. In this case, the distribution of logarithmic conductance is usually analyzed instead.\cite{Melnikov,Abrikosov} Fig.~\ref{fig:1D-G}(d) shows a one-sided log-normal distribution in the diffusive regime at $W=2.76$, and a log-normal distribution is found at larger disorder $W=5$ in Fig.~\ref{fig:1D-G}(e). Recall that in 1D normal metal systems, Gaussian, one-sided log-normal,\cite{Dorokhov,Mello,Plerou,Muttalib,Garcia} and log-normal distributions\cite{Janssen,Plerou,Muttalib,Ren1} were found in the metallic, diffusive, and localized regimes, respectively. Here, the universal exponent $n=1$ together with the log-normal distribution of ${\rm ln}G_{T}$ suggests that SPS also works. Therefore, statistical properties of the magnon-mediated spin transport in 1D disordered NM/FI systems is consistent with that of charge transport in 1D normal metal systems.

\subsection{Disorder-enhanced spin transport in 2D NM/FI system}\label{sec:enhance}

\begin{figure}
\includegraphics[width=8.5cm]{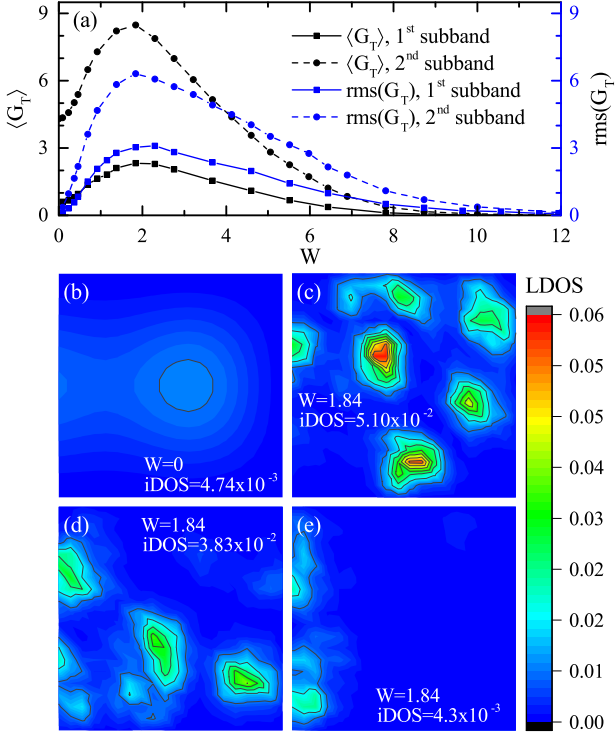}%
\caption{\label{fig:G} (a) Average $\langle G_{T}\rangle$ and its fluctuation $\mathrm{rms}(G_{T})$ as a function of $W$ in the first and second subbands of the 2D NM/FI system. Parameters for the first and second subbands: $\mu_{L}=\{0.9, 3\}$ meV, $\omega_{c}=\{0.24,0.2\}$ meV, and $T=5$ K. (b) Local DOS in the scattering region without disorder. (c)--(e) LDOS for three typical configurations at disorder strength $W=1.84$. The spin conductances for these configurations are $G_T=\{4.107, 4.7, 0.0786\}$, respectively. Here iDOS is the total interfacial DOS at the NM/FI interface and we set parameters in the first subband.}
\end{figure}

In 2D calculations, we consider a central region with size $20\times 20$.\cite{note3} One left NM lead and one right FI lead with the same width are attached to it, as shown in Fig.~\ref{fig:setup}. To reduce computational cost, we focus on the first and second electronic subbands of the left lead. The effective coupling strength $\alpha$ is chosen to be $\alpha=10^{3}$, which is often used in the interface of Pt/YIG materials.\cite{Zheng} The cutoff frequency for the first and second subbands are $\omega_{c}=0.24$ meV and $0.2$ meV, so that the magnonic spectra decays nearly to zero at the subband edges.

Fig.~\ref{fig:G}(a) depicts average spin conductance $\langle G_T\rangle$ and its fluctuation $\mathrm{rms}(G_T)$ for the first and second subbands of the 2D NM/FI system. It shows that $\langle G_T\rangle$ and $\mathrm{rms}(G_T)$ are also in the same order of magnitude. In 2D normal metal systems, disorder is known to suppress electronic transport monotonically. In Fig.~\ref{fig:G}(a), average spin conductance varies non-monotonically with increasing $W$. In the first subband, $G_{T}=0.6$ is found for the clean system with $W=0$. When weak disorder is present, $\langle G_T\rangle$ is enhanced. The largest enhancement, about $5$ times, is reached around $W=1.84$ with $\langle G_T\rangle=2.33$, accompanied by the largest fluctuation $\mathrm{rms}(G_T)$. Further increasing of disorder continuously suppresses the average spin conductance till spin transport is completely blocked beyond $W=12$. Similar behaviors are also for the second subband, except a smaller enhancement, which indicates that the spin conductance enhancement is a general property.

These unusual disorder-enhancement behaviors in contrast to Anderson localization suggest that there is another mechanism dominating spin transport in the weak disorder regime. From Eq.~(\ref{ar}), it is clear that the quantity $A_{R}$ and hence the spin conductance are determined by local DOS matrices $D_{L\uparrow}(E)$ and $D_{L\downarrow}^{0}(\bar{E})$. $D_{L\uparrow}(E)$ contains LDOS information of the whole system, which is chosen for demonstration. For a clean system, its LDOS landscape in Fig.~\ref{fig:G}(b) is smooth and $G_T=0.5$. In the presence of disorder, local potential $U = [-W/2, W/2]$ can be negative, which will energetically increase the electron dwell time or LDOS. Hence there are many peaks in the LDOS landscape, as shown in Fig.~\ref{fig:G}(c)--(e) for $W = 1.84$. The corresponding spin conductances are $G_T=\{4.107, 4.70, 0.0786\}$, respectively. There are both enhancement and suppression at the same disorder. When the LDOS peaks are near the NM/FI interface, the interfacial DOS (iDOS) increases (labeled in Fig.~\ref{fig:G}(b)--(e)). Comparing with the clean system in Fig.~\ref{fig:G}(b), the iDOS is increased by a factor of 11 or 8 in Fig.~\ref{fig:G}(c)--(d), respectively, which is responsible for the enhancement of spin conductance. For the disorder configuration in Fig.~\ref{fig:G}(e), there are few LDOS peaks which are far away from the interface, resulting in a small iDOS. The spin conductance is thus suppressed. Therefore, the disorder-enhanced iDOS causes the enhancement of spin conductance. For strong disorder, electron wave function is more localized and the LDOS peaks are much sharper. Besides, it is more difficult for electrons to reach the NM/FI interface in strong disorders. As a result, the iDOS decreases drastically and enhancement vanishes.

Notice that the enhancement of spin transport occurs only in 2D systems. In 1D systems, there is only one interface site where the random potential is either positive or negative with equal probability. Since positive (negative) potential decreases (increases) the LDOS, on average there is no effect on the spin transport in 1D systems. For 2D systems, however, in any configuration, there are always some sites with negative potentials at the NM/FI interface so that electrons can dwell around these sites leading to the enhancement of iDOS and hence the enhancement of spin conductance in weak disorders. While further increasing disorder strength ($W>2$), the localized state induced by strong disorders dominates, which suppresses average spin conductance. These two competing contributions lead to the non-monotonic line shape in average spin conductance. Our results are consistent with the previous study on this NM/FI system, which achieved an enhancement of nearly three orders of magnitude on the spin conductance via engineering the interfacial potentials.\cite{Li}

\begin{figure}[tbp]
\includegraphics[width=8.cm]{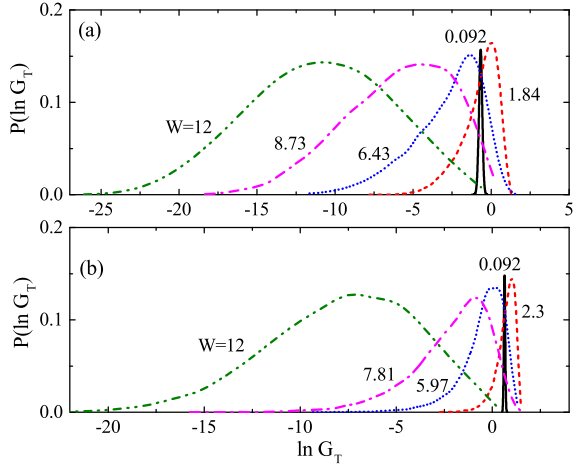}%
\caption{\label{fig:dist}Spin conductance distribution $P(\mathrm{ln}G_{T})$ of the 2D system for different $W$ in the first subband (a) and the second subband (b). Parameters are the same as Fig.~\ref{fig:G}.}
\end{figure}

\subsection{Statistical behaviors in 2D NM/FI system}

We first investigate the spin conductance distribution in 2D NM/FI systems. Fig.~\ref{fig:dist} shows distributions for different disorder strengths in the first and second subbands. In the first subband, for weak disorder ($W = 0.092$), $P(G_T)$ is Gaussian like. For intermediate disorders $W=\{1.84, 6.43, 8.73\}$, the distribution $P(G_T)$ follows the one-sided log-normal shape. While in the strong disorder region ($W\simeq 12$), $P({\rm ln}G_T)$ is nearly Gaussian. The distribution in the second subband is quite similar to that of the first subband. The peak positions of lines $W=1.84$ and $W=2.3$ in Fig.~\ref{fig:dist} (a) and (b) lie to the right of line $W=0.092$. This corresponds to the weak disorder regime where average spin conductance is enhanced ($W<2$ in Fig.~\ref{fig:G}). Expect this difference, these findings are similar to the results of 2D normal systems,\cite{Ren1} but the scaling behaviors are quite different.

\begin{figure}[tbp]
\includegraphics[width=\columnwidth]{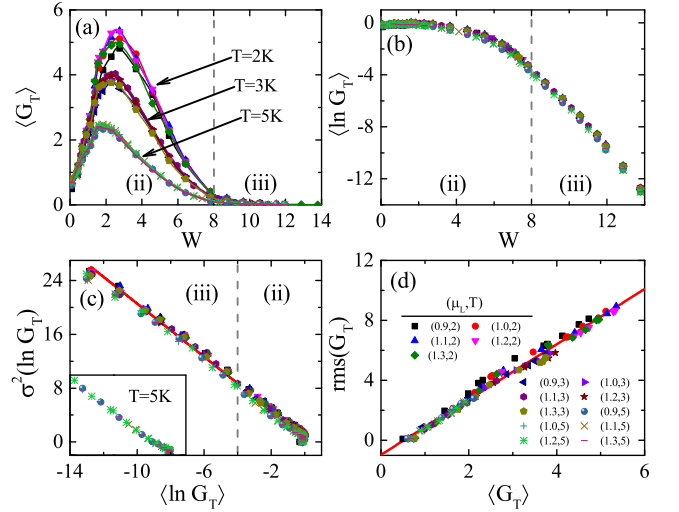}
\caption{\label{fig:scaling}Statistics of spin conductance for Fermi energies $\mu_{L}=\{0.9, 1.0, 1.1, 1.2, 1.3\}$ meV and temperatures $T=\{2,3,5\}$ K. All Fermi energies are in the first subband. Figure legends are shown in (d). The gray dashed lines separate (ii) the diffusive regime and (iii) the localized regime. (a) Average spin conductance as a function of the disorder strength $W$. (b) $\langle{\rm ln} G_{T}\rangle$ versus the disorder strength. (c) Scaling of the variance $\sigma^{2}({\rm ln}G_{T})$ on average ${\rm ln}G_{T}$ in the localized regime. Inset: data for different Fermi energies at $T=5$ K. (d) Scaling of spin conductance fluctuation on average spin conductance in the metallic regime.}
\end{figure}

We next study the scaling properties of spin transport in 2D NM/FI systems. We focus on the first subband. The average spin conductance for different Fermi energies and temperatures are plotted in Fig.~\ref{fig:scaling}(a). Disorder-enhancement of spin conductance is observed in weak disorder for all parameters. At lower temperature, the system has larger average spin conductance and stronger fluctuation. Raising temperature suppresses the average spin conductance and its fluctuation. Fig.~\ref{fig:scaling}(b) shows the average logarithm spin conductance as a function of the disorder strength. It is interesting that $\langle{\rm ln} G_{T}\rangle$ falls into a single curve for different Fermi energies and temperatures, indicating the existence of universal scaling.

Previous study on charge transport in 2D normal metal systems has shown that, in the localized regime, the variance of charge conductance $\sigma^{2}(\ln G)$ scales as $\sigma^{2}(\ln G) \propto \langle{\rm ln}G\rangle^{2/3}$.\cite{Somoza} For the 2D NM/FI system, the scaling relation is shown in Fig.~\ref{fig:scaling}(c). We see that data points for $\langle{\rm ln} G_{T}\rangle<-4$, or equivalently $W>8$, collapse into a universal line (red line) well-fitted by
\begin{equation}\label{eq:scaling2}
\sigma^{2}({\rm ln}G_T) = -1.98 \langle{\rm ln}G_{T}\rangle + B.
\end{equation}
The intercept $B$ depends weakly on Fermi energies and temperatures. The inset depicts the variance dependence on $\langle{\rm ln} G_{T}\rangle$ at $T=5$ K, which shows better linear behavior. We conclude that for 2D NM/FI systems in the localized regime, universal scaling relation Eq.(\ref{eq:scaling}) is justified with $n=1$ and the scaling weakly depends on temperature.

For the region out of the localized regime, it is difficult to study the scaling properties using ${\rm ln}G_{T}$ since data points are scattered for large $\langle{\rm ln}G_{T}\rangle$. However, using $\langle G_{T}\rangle$ and its fluctuation ${\rm rms}(G_{T})$, we find an additional scaling behavior at weak disorders. The ${\rm rms}(G_{T})$ dependence on $\langle G_{T}\rangle$  for $W<2$ is plotted in Fig.~\ref{fig:scaling}(d), which are well fitted to a straight line
\begin{equation}
{\rm rms}(G_{T}) = 1.84 \langle G_{T}\rangle - 0.96.
\end{equation}
Note that at weak disorder, the charge (spin) conductance fluctuation always increases with increasing of disorder since there is no fluctuation when disorder is zero. The positive slope indicates that the average spin conductance increases with disorder strength, i.e., enhancement of spin conductance at weak disorders.

As discussed above, our calculation enables us to verify three regimes in this 2D NM/FI system. (i) The metallic regime for $W<2$, where the average spin conductance is enhanced due to the large DOS at the NM/FI interface when increasing disorder. The fluctuation ${\rm rms}(G_{T})$ scales linearly with $\langle G_{T}\rangle$. (ii) The diffusive regime for $2<W<8$, where localization caused by disorders in the entire scattering region strongly competes with interfacial resonance, leading to a non-monotonic average spin conductance. (iii) The localized regime for $W>8$, in which Anderson localization dominates. The variance $\sigma^{2}(\ln G_T)$ for different disorder strengths, Fermi energies can be linearly scaled by $\langle \ln G_{T}\rangle$. The three regimes are labeled in Fig.~\ref{fig:scaling} and separated by gray dashed lines. Calculations in the second subband show similar results.

\begin{figure}
\includegraphics[width=8.5cm]{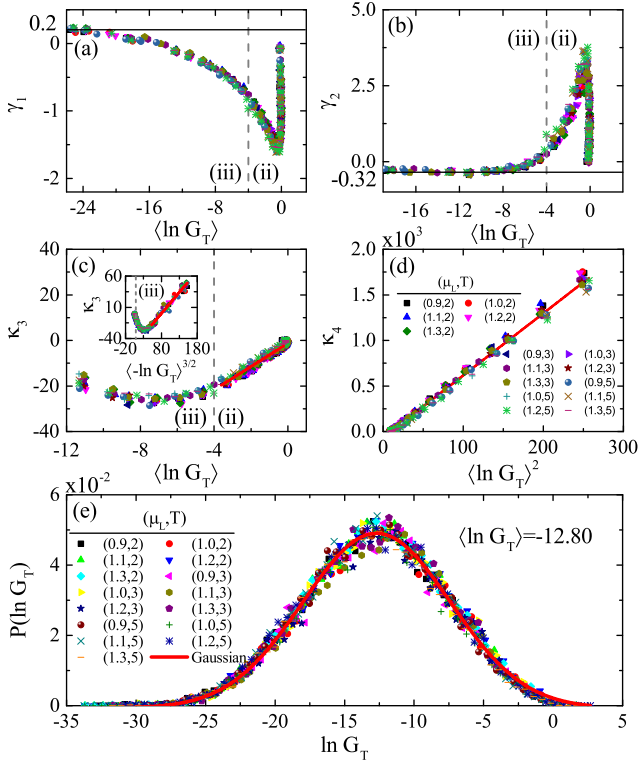}%
\caption{\label{fig:k3k4} (a)--(d) Skewness $\gamma_{1}$, kurtosis $\gamma_{2}$, third and fourth cumulants $\kappa_{3}, \kappa_{4}$ of ${\rm ln}G_T$ for the 2D NM/FI system in the first subband. Parameters are the same as in Fig.~\ref{fig:scaling}.\cite{note4} Legends are shown in (d). The gray dashed lines separate the localized regime and diffusive regime. The inset of (c) shows $\kappa_3$ versus $\langle -\ln G_{T}\rangle^{3/2}$. (e) $P(\ln G_T)$ at a fixed value $\langle {\rm ln} G_{T}\rangle=-12.80$ for different Fermi energies and temperatures in the localized regime with $W = 13.78$. The red curve is the fitted Gaussian distribution.}
\end{figure}

Since $P(\ln G_T)$ in the localized regime is nearly Gaussian in Fig.~\ref{fig:dist}, we examine its deviation from Gaussian distribution by calculating the skewness, kurtosis and the lowest nontrivial cumulants of ${\rm ln} G_{T}$. The definitions of the third and fourth cumulants $\kappa_{3}, \kappa_{4}$, skewness $\gamma_{1}$ and kurtosis $\gamma_{2}$ used here are the same as those defined in Refs.~[\onlinecite{Ren1}] and [\onlinecite{Mohanty}] with $\kappa_{3}=\mu_{3}, \kappa_{4}=\mu_{4}$, and
\begin{eqnarray}
\gamma_{1} = \frac{\mu_{3}}{\mu_{2}^{3/2}},\\
\gamma_{2} = \frac{\mu_{4}}{\mu_{2}^{2}} - 3,
\end{eqnarray}
where $\mu_{m}=\langle(x-\langle x\rangle)^{m} \rangle$ is the m-$th$ order central moment. Skewness is usually used to quantify the asymmetry of a distribution. It is positive when the distribution has a long flat tail in larger values, while a zero skewness indicates that the distribution is symmetric about its mean value. Kurtosis is a measure of sharpness or flatness of a distribution. It is zero for a Gaussian distribution, greater than zero if the distribution has a sharper peak compared to a normal distribution, and less than zero if the distribution peak is flatter.

The skewness, kurtosis, third and fourth cumulants of ${\rm ln}G_T$ are depicted in Fig.~\ref{fig:k3k4}(a)--(d) for different Fermi energies and temperatures in the first subband. They show good universal behaviors. In the localized regime, the skewness and kurtosis approach to nonzero values $0.2$ and $-0.32$, respectively [the horizontal lines in Fig.~\ref{fig:k3k4}(a) and (b)]. Thus the ${\rm ln} G_{T}$ distribution in this regime is not precisely Gaussian. The third cumulant in Fig.~\ref{fig:k3k4}(c) shows a good linear dependence on $\langle {\rm ln}G_T\rangle$ in the diffusive regime, which is denoted by the red line fitted as $\kappa_3=5.11\langle {\rm ln}G_T\rangle - 1.26$. But in the localized regime, $\kappa_{3}$ is approximately scaled as $\langle -{\rm ln} G_{T}\rangle^{3/2}$ [see inset of Fig.~\ref{fig:k3k4}(c)]. In addition, the fourth cumulant $\kappa_{4}$ scales linearly with $\langle {\rm ln}G_{T}\rangle^{2}$ in all three regimes and shows weak dependence on temperature (Fig.~\ref{fig:k3k4}(d)),
\begin{equation}
\kappa_{4} =6.81\langle {\rm ln}G_{T}\rangle^{2} - 65.46.
\end{equation}
In Fig.~\ref{fig:k3k4}(d), we use meVK$^{-1}$ as the unit of spin conductance. In contrast, universal behaviors of higher order moments or cumulants (third and fourth) for charge transport in localized 2D normal metal systems have been reported as: $\kappa_3 \sim \langle -\ln G \rangle$ and $\kappa_4 \sim \langle -\ln G \rangle^{4/3}$.\cite{Somoza1}

From the universal behaviors of moments of $\langle \ln G_T \rangle$ in the localized regime, we anticipate the distribution of spin conductance may also be universal in the same regime. Indeed, as shown in Fig.~\ref{fig:k3k4}(e), different data with the same $\langle {\rm ln}G_{T}\rangle$ approximately collapse into a single curve, suggesting that the distribution $P(\ln G_T, \langle {\rm ln}G_{T}\rangle)$ is a universal function. The curve is fitted by a Gaussian function, where the deviation is accounted for nonzero higher order cumulants.

\subsection{Finite size effects in 1D and 2D NM/FI systems}

It is important to inspect the statistical behavior of spin conductance while increasing the system size. It is well known that in normal metal systems, the charge conductance decays exponentially as the system size increases in the localized regime, which is a strong evidence for Anderson localization.

\begin{figure}
\includegraphics[width=\columnwidth]{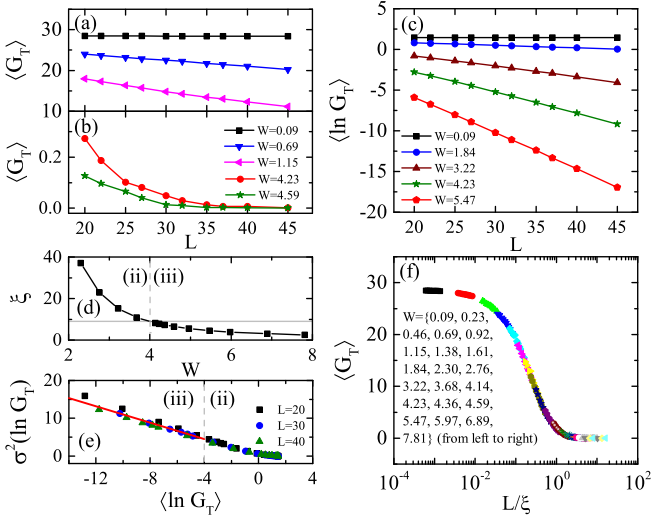}%
\caption{\label{fig:1Dsize} $\langle G_T \rangle$ [(a) and (b)] and $\langle {\rm ln}G_T \rangle$ (c) versus the 1D system length $L$ for different $W$. (d) Localization length $\xi$ extracted from the slopes in (c). (e) Universal scaling of $\sigma^2({\rm ln}G_T)$ on $\langle {\rm ln} G_T\rangle$ in the localized regime (iii) for different system lengths. The red line is the same as in Fig~\ref{fig:1D-G}(c). (f) The scaling of $\langle G_T\rangle$ as a function of the ratio $L/\xi$ for different disorder strengths. 30000 disorder samples are collected to smooth the curve. The gray dashed line corresponds to the localization length $\xi\approx 10$, which is much smaller than the system size.}
\end{figure}

For 1D NM/FI systems, we show in Fig.~\ref{fig:1Dsize} the finite size scaling calculation. The average spin conductance $\langle G_T\rangle$ of 1D chains with lengths $L=\{20, 22, 25, 27, 30, 32, 35, 37, 40, 45\}$ is calculated for a large range of disorder strength. At weak disorders [Fig.~\ref{fig:1Dsize}(a)], $\langle G_T \rangle$ decays linearly with the system length. At strong disorders [Fig.~\ref{fig:1Dsize}(b)], $\langle G_T \rangle$ decays in a nearly exponential form as $L$ increases. In Fig.~\ref{fig:1Dsize}(c), we plot $\langle {\rm ln}G_T\rangle$ versus the system length $L$. $\langle {\rm ln}G_T\rangle$ shows good linear dependence on the system length for large $W$, which corresponds to Anderson localization. This allows us to extract the localization length $\xi$ from the linear fit
\begin{equation}
\langle {\rm ln} G_T \rangle = -\frac{2}{\xi}L + C.
\end{equation}
The localized regime is reached when $\xi\ll L$. In the NM/FI hybrid system, spin conductance in the localized regime is suppressed to nearly zero since the electron transport is localized in the disordered central region. In Fig.~\ref{fig:1Dsize}(d), we plot the extracted $\xi$. This result is consistent with the partition in Fig.~\ref{fig:1D-G}, where the localized regime locates in the range $W>4$. The localization length $\xi\approx 10$ is much smaller than the system length $L=30$ used in Fig.~\ref{fig:1D-G}. In the localized regime, Fig.~\ref{fig:1Dsize}(e) shows that the scaling of $\sigma^2({\rm ln}G_T)$ on $\langle {\rm ln} G_T\rangle$ remains universal for different system lengths. The red line denotes the same one in Fig~\ref{fig:1D-G}(c). Fig.~\ref{fig:1Dsize}(f) shows that the average spin conductance $\langle G_T \rangle$ depends only on one parameter $L/\xi$, in good agreement with the single-parameter scaling theory.

\begin{figure}
\includegraphics[width=\columnwidth]{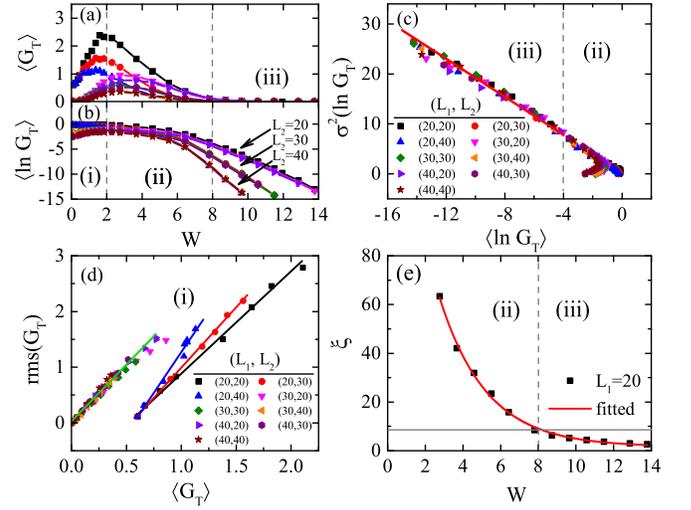}%
\caption{\label{fig:size} The average of $G_T$ (a), ${\rm ln}G_T$ (b) and the scaling in the localized regime (c) for different 2D system widths $L_1=\{20, 30, 40\}$ and lengths $L_2=\{20, 30, 40\}$. The legends for (a)--(c) is shown in (c). (d) The scaling of ${\rm rms}(G_T)$ on $\langle G_T \rangle$ in the metallic regime for different $L_1$ and $L_2$. (e) The localization length $\xi$ of system width $L_1=20$ as a function of $W$. The black points are numerically extracted and the red curve is the fitted exponential function. Parameters are chosen in the first subband, $\mu_L=\{0.9, 0.5, 0.3\}$ meV, $T=\{5, 2, 1.1\}$ K, and $\omega_c=\{0.24, 0.05, 0.03\}$ meV for width $L_1=\{20, 30, 40\}$, respectively.}
\end{figure}

For 2D NM/FI systems, we calculate the spin conductance $G_T$ for different system widths $L_1$ and lengths $L_2$. As a comparison to Fig.~\ref{fig:scaling}, Fig.~\ref{fig:size}(a)--(d) show the same quantities. Fig.~\ref{fig:size}(a) demonstrates that disorder-enhancement of spin conductance is a general feature due to the competition between bulk and interface physics. In Fig.~\ref{fig:scaling}(b), $\langle{\rm ln} G_T\rangle$ for different Fermi energies and temperatures collapse with each other in the whole range of disorder. While varying system width and length as shown in Fig.~\ref{fig:size}(b), the scaling of $\langle{\rm ln} G_T\rangle$ with respect to $W$ still exists. It is independent of system width but depends on system length. Scaling of ${\rm ln} G_T$ for different system widths and lengths is shown in Fig.~\ref{fig:size}(c). For the system size $20\times 20$, we use the same data as in Fig.~\ref{fig:scaling}(c) with $\mu_L=0.9$ meV and $T=5$ K. Clearly, the scaling of variance $\sigma^2({\rm ln} G_T)$ on $\langle {\rm ln} G_T \rangle$ is still universal in the localized regime. The variance $\sigma^2({\rm ln} G_T)$ follows the same equation [Eq.~(\ref{eq:scaling2})] in a large range, from diffusive to localized regimes, which is denoted by the red line in Fig.~\ref{fig:size}(c). In the metallic regime [Fig.~\ref{fig:size}(d)], the linear relation rms$(G_T) \propto \langle G_T\rangle$ remains. But the proportionality constant depends on the system size and fluctuates heavily for small width $L_1=20$. While for larger widths 30 and 40, the dependence of rms($G_T$) on $\langle G_T\rangle$ tends to be universal, which is well fitted by rms$(G_T)=2.04\langle G_T\rangle+0.03$ and denoted by the green line.

The size dependence of spin conductance in 2D NM/FI systems is similar to the 1D case. $\langle G_T \rangle$ decays linearly at weak disorders and exponentially at strong disorders. In Fig.~\ref{fig:size}(e), we plot the localization length $\xi$ as a function of disorder strength $W$ for a fixed width $L_1=20$. Then $\langle {\rm ln}G_T\rangle$ depends linearly on the system length $L_2$: $\langle {\rm ln} G_T \rangle = - 2 L_2 / \xi + C$. The localization length is extracted when varying the length $L_2 =\{20, 25, 30, 35, 40\}$. The dependence of $\xi$ on $W$ is well fitted by an exponential decay function: $\xi(W) = 1.6 + 184.8e^{-0.4 W}$, which is denoted by the red curve.
The dashed vertical line indicates the critical disorder strength above which the localized regime is reached. We can see that the localization length at $W=8$ is less than 10, which is much smaller than the system size $20 \times 20$ used in Fig.~\ref{fig:scaling}.

\section{Conclusion}\label{sec:conclusion}

In summary, we have numerically investigated magnon-mediated spin transport in disordered 1D and 2D NM/FI heterostructures based on the NEGF method. In 1D, disorder suppresses the spin conductance $G_T$. The distribution of $G_T$ is in good agreement with that of charge transport in 1D normal metal systems. In 2D NM/FI systems, average spin conductance is enhanced at weak disorder and suppressed at strong disorder, which is attributed to the competition between resonance-induced increasing of interfacial DOS at the NM/FI interface and electron localization in the central region.

Universal behaviors of spin conductance are discovered. For both 1D and 2D NM/FI systems, the variance $\sigma^2({\rm ln}G_{T})$ varies linearly with average $\langle {\rm ln}G_{T}\rangle$ in the localized regime, which is universal and independent of parameters such as Fermi energy, temperature, and system size. The linear scaling of $\langle {\rm ln}G_{T}\rangle$ in 2D is different from the $2/3$ power-law of charge transport in 2D normal metal systems.\cite{Prior} In the localized regime, the third and fourth order cumulants of ${\rm ln}G_{T}$ exhibit universal behaviors, which are $\kappa_3 \sim \langle -\ln G_{T}\rangle^{3/2}$ and $\kappa_4 \sim \langle -\ln G_{T}\rangle^{2}$, respectively; the distribution of ${\rm ln}G_{T}$ is approximately Gaussian and depends only on $\langle \ln G_{T}\rangle$. Moreover, in the metallic regime of 2D systems, spin conductance fluctuation ${\rm rms}(G_{T})$ scales linearly with its average $\langle G_{T}\rangle$, independent of system parameters for large system width. These results reveal that magnon-mediated spin transport in disordered 2D systems belongs to a new universality class different from that of charge transport in normal metal systems.

\section*{acknowledgments}

This work was supported by the Natural Science Foundation of China (Grants No.12034014 and No.12174262).


\begin{thebibliography}{00}
\bibitem{note2} In Ref.~[\onlinecite{Prior}], ${\rm ln}R$ is used instead of ${\rm ln}G$. Note the resistence $R$ is the inverse of conductance $G$, then we have $\langle{\rm ln}G\rangle=-\langle{\rm ln}R\rangle$ thus the same result is obtained.
\bibitem{note3} When calculating spin conductance for one specific disorder sample, one has to perform a double integration (see Eq.(\ref{spin4})). In the calculation, we have chosen grid points $100 \times 100$ which amounts to an increase of computational cost by the factor of $10000$ comparing with the charge conductance calculation in normal metal systems.
\bibitem{note4} The Fermi energies used in scaling analysis (Fig.~\ref{fig:scaling} and Fig.~\ref{fig:k3k4}) are near the center of the first subband. This is the region where SPS works in normal metal system.\cite{Prior} Although the ranges of Fermi energies and temperatures are not very wide, the spin conductance in Fig.~\ref{fig:scaling}(a) varies large enough in magnitude to support our scaling analysis.
\end{thebibliography}
\end{document}